\documentclass{ws-ijmpcs}

\usepackage{float}

\begin{document}


\markboth{Masayasu Harada and Yong-Liang Ma} {Properties of
$X(3872)$ as a hadronic molecule with a negative parity}

%
\catchline{}{}{}{}{}
%

\title{Properties of $X(3872)$ as a hadronic molecule with a negative parity
}

\author{Masayasu Harada
}

\address{Department of Physics, Nagoya University, Nagoya, 464-8602, Japan.
\\
harada@hken.phys.nagoya-u.ac.jp}

\author{Yong-Liang Ma}

\address{Department of Physics, Nagoya University, Nagoya, 464-8602, Japan.\\
ylma@hken.phys.nagoya-u.ac.jp}

\maketitle



\begin{abstract}
We discuss the possible interpretation of $X(3872)$ as a $DD^{\ast}$
hadronic molecule with $J^{PC} = 2^{-+}$. Using the phenomenological
Lagrangian approach, we studied its radiative and strong decay
properties. We find that our model with about 97.6\% isospin zero
component explains the existing data nicely. We predict the partial
widths of the radiative and strong decays of $X(3872)$ and show that
the measurement of the ratio $\mathcal{B}(X(3872) \to
\chi_{c0}\pi^0)/\mathcal{B}(X(3872) \to \chi_{c1}\pi^0)$ may signal
the nature of $X(3872)$.

\keywords{Exotic state; hadronic molecule; negative parity.}

\end{abstract}

\ccode{14.40.Rt, 13.20.Fc, 13.25.Ft.}


\section{Introduction}

In our recent work~\cite{Harada:2010bs}, we presented a composite
model for $X(3872)$ with $J^{PC} = 2^{-+}$ which is favored by
recent experiment~\cite{al.:2010jr} by regarding it as a $DD^\ast$
bound state. Based on an effective Lagrangian approach, we fitted
the parameters using the existing data and found that $X(3872)$ is
dominantly an isospin singlet. With this clear and straight model,
we calculated the radiative and strong decay properties of
$X(3872)$.

\section{Hadronic Molecular Structure of X(3872) with $J^{PC} = 2^{-+}$}

We write the wave function of $X(3872)$ in terms of charge
eigenstates as
\begin{eqnarray}
|X(3872)\rangle = \frac{\cos\theta}{\sqrt{2}}| D^{0}\bar{D}^{\ast \,
0}\rangle + \frac{\sin\theta}{\sqrt{2}}| D^{+}D^{\ast \, -}\rangle +
{\rm C.c.},\label{eq:defmix}
\end{eqnarray}
or equivalently, in terms of the isospin eigenstates as
\begin{eqnarray}
|X(3872)\rangle = \cos\phi|X(3872)\rangle_{I=0} +
\sin\phi|X(3872)\rangle_{I=1}, \label{eq:mixiso}
\end{eqnarray}
with $\cos\theta = (\cos\phi + \sin\phi)/\sqrt{2}$ and $\sin\theta =
(\cos\phi - \sin\phi)/\sqrt{2}$ and
\begin{eqnarray}
|X(3872)\rangle_{I=0,1} & = & \frac{1}{2}\Big(| D^{0}\bar{D}^{\ast
\, 0}\rangle \pm | D^{+}D^{\ast \, -}\rangle \Big) + {\rm
C.c.}, 
\end{eqnarray}

Our composite model is based on the effective Lagrangian describing
the interaction between $X(3872)$ and its constituents
\begin{eqnarray}
{\cal L}_{\rm X} & = & \frac{i}{\sqrt{2}}X^{\mu\nu}(x)\int dx_1
dx_2 \Phi_X((x_1-x_2)^2)\delta(x-\omega_v x_1 - \omega_px_2)\nonumber\\
& & \times \Big\{g_{_X}^{N}\Big[ C_{\mu\nu}^{N}(x_1,x_2) +
C_{\nu\mu}^{N}(x_1,x_2) -
\frac{1}{4}g_{\mu\nu}C_{\alpha}^{N;\alpha}(x_1,x_2)\Big]
\nonumber\\
& & \;\;\;\; + g_{_X}^{C}\Big[ C_{\mu\nu}^{C}(x_1,x_2) +
C_{\nu\mu}^{C}(x_1,x_2) -
\frac{1}{4}g_{\mu\nu}C_{\alpha}^{C;\alpha}(x_1,x_2)\Big]
\Big\},\label{effelcomp}
\end{eqnarray}
where $g_{_X}^{N} (g_{_X}^{C})$ is the effective coupling constant
for the interaction between $X(3872)$ and its neutral (charged)
constituents. $\omega_v$ and $\omega_p$ are mass ratios with
definitions $ \omega_v = m_{_{D^{\ast}}}/(m_{_{D^{\ast}}} +
m_{_{D}}), \omega_p = m_{_{D}}/(m_{_{D^{\ast}}} + m_{_{D}})$ with
$m_{_D}(m_{_{D^{\ast}}})$ as the mass of the constituent
$D(D^{\ast})$ meson. The function $\Phi_X((x_1-x_2)^2)$ illustrates
the finite size of the molecule, and in the calculation, its Fourier
transform $\tilde{\Phi}_X(p^2) =
 \exp(p^2/\Lambda_{_X}^2)$
with the size parameter $\Lambda_{_X}$ parameterizing the
distribution of the constituents inside the molecule has been
applied. The tensor $C_{\mu\nu}^{N}$ is defined as
\begin{eqnarray}
C_{\mu\nu}^{N}(x_1,x_2) & = & \bar{D}_{\mu}^{\ast \,
0}(x_1)\partial_\nu D^{0}(x_2) + D_{\nu}^{\ast \,
0}(x_1)\partial_\mu \bar{D}^{0}(x_2) .
\end{eqnarray}
Substituting the neutral constituents with the corresponding charged
ones, one can get the explicit form of $C_{\mu\nu}^{C}$.

Relation between the mixing angle $\theta$ defined in
Eq.~(\ref{eq:defmix}) and the coupling constant
$g_{_X}^{N}(g_{_X}^{C})$ can be yielded using the compositeness
condition $Z_X = 0$~\cite{Weinberg:1962hj,Salam:1962ap} with $Z_X$
as the wave function renormalization constant of $X(3872)$ which is
defined as
\begin{eqnarray}
Z_X & = & 1 -
g_{_X}^2\frac{d}{dp^2}\Sigma_{_X}(p^2)\Big|_{p^2=m_{_X}^2} ,
\end{eqnarray}
where $g_{_X}^2\Sigma_{_X}(p^2)$ relates to the mass operator via
the relation
\begin{eqnarray}
\Pi_{_X}^{\mu\nu;\alpha\beta}(p^2) & = &
\frac{1}{2}(g^{\mu\alpha}g^{\nu\beta} +
g^{\mu\beta}g^{\nu\alpha})g_{_X}^2\Sigma_{_X}(p^2) + \cdots ,
\end{eqnarray}
with $``\cdots"$ denoting terms do not contribute to the mass
renormalization of $X(3872)$. Explicitly, we have the following
relations
\begin{eqnarray}
\cos\theta & = &
g_{_X}^{N}\frac{d}{dp^2}\Sigma_{_X}^{N}(p^2)\Big|_{p^2 =
m_{_X}^2},\;\;\;\;\; \sin\theta =
g_{_X}^{C}\frac{d}{dp^2}\Sigma_{_X}^{C}(p^2)\Big|_{p^2 = m_{_X}^2},
\end{eqnarray}
where $\Sigma_{_X}^{N} (\Sigma_{_X}^{C})$ corresponds to the case
with neutral (charged) constituents.

Concerning the uncertainty of the $X(3872)$ mass, we express $m_{_X}
= m_{D^{\ast \, 0}} + m_{D^0} - \Delta E$ with $\Delta E > 0$ as the
binding energy and $m_{D^{\ast \, 0}} = 2006.97~$MeV and $m_{D^0} =
1864.84~$MeV~\cite{Amsler:2008zz}. In the computation, we take
$\Delta E = 0.5, 1.0$ and $1.5~$MeV.

\section{Radiative and Strong Decay Properties}

Using the model (\ref{effelcomp}), we calculated the radiative and
strong decay widths. Except the coupling constant between the
molecule and its constituents which is determined using the
compositeness condition, other coupling constants are borrowed from
the effective Lagrangian approaches.

In our numerical calculation, we take the mixing angle $\phi$ and
the size parameter $\Lambda_{_X}$ as free parameters to fit the
following data
\begin{eqnarray}
R_1 & = & \frac{\mathcal{B}(X \to \gamma \psi(2S))}{\mathcal{B}(X
\to \gamma J/\psi)} = 3.4 \pm 1.4 ~ \mbox{BaBar~\cite{:2008rn}},\label{dataratioem}\\
R_2 & = & \frac{\mathcal{B}(X(3872) \to
J/\psi\pi^+\pi^-\pi^0)}{\mathcal{B}(X(3872)\to J/\psi\pi^+\pi^-)} =
1.0 \pm 0.4({\rm stat.}) \pm 0.3 ({\rm syst.})~\mbox{Belle~\cite{Abe:2005ix}}, \label{eq:data3pto2p}\\
R_3 & = & \frac{\mathcal{B}(X(3872) \to \gamma
J/\psi)}{\mathcal{B}(X(3872) \to J/\psi\pi^+\pi^-)} = 0.14 \pm
0.05~\mbox{Belle~\cite{Abe:2005ix}}; \; 0.33 \pm 0.12
~\mbox{BaBar~\cite{:2008rn}}.\label{gatagamma2pi}
\end{eqnarray}
We find that we can reproduce all the above data except $R_3$ by
Belle~\cite{Abe:2005ix}. If the Belle data for $R_3$ is favored in
future experiments, our model will be excluded. It should be
stressed that we can reproduce the above three data with only two
parameters so our model is nontrivial. Combining with the data $R_4
= \mathcal{B}(X(3872) \to \gamma \psi(2S))/\mathcal{B}(X(3872) \to
J/\psi \pi^+\pi^-) = 1.1 \pm 0.4$ from BaBar~\cite{:2008rn}, we
restrict ourselves to the case with $\phi = 9^\circ$ and the
corresponding $\Lambda_{_X} = 2.8 - 3.0~$GeV. Our results of
relevant ratios are given in Table.~\ref{table:fit}.
\begin{table}[h]
\tbl{The ratio with fitted parameters $\phi = 9^\circ$ and
$\Lambda_{_X} = 2.8 - 3.0~$GeV. The last two rows are data.}
{\begin{tabular}{lllll} \hline \hline $\Delta E$(MeV) & $R_1$ &
$R_2$ & $R_3$  & $R_4$ \\
\hline
$ 0.5$ & $2.430 - 3.486$ & $0.964 - 1.035$ & $0.369 - 0.283$ & $0.897 - 0.987$ \\
$ 1.0 $ & $2.433 - 3.504$ & $0.917 - 0.971$ & $0.363 - 0.274$ & $0.883 - 0.960$ \\
$ 1.5$ & $2.433 - 3.516$ & $0.874 - 0.914$ & $0.358 - 0.266$ & $0.871 - 0.935$ \\
\hline
~ & $3.4 \pm 1.4$~\cite{:2008rn} & $1.0 \pm 0.4 \pm 0.3 $~\cite{Abe:2005ix} & $0.14 \pm 0.05 $~\cite{Abe:2005ix} & $1.1 \pm 0.4$~\cite{:2008rn} \\
~ & ~ & ~ & $0.33 \pm 0.12 $~\cite{:2008rn} & ~\\
\hline \hline
\end{tabular}
\label{table:fit}}
\end{table}
\begin{table}[h]
\tbl{Effective coupling constants $g_{_X}^{N}$ and $g_{_X}^{C}$ from
the fitted parameters $\phi$ and $\Lambda_{_X}$.}
{\begin{tabular}{lllll} \hline \hline \,\, $\Delta E$(MeV) \,\, &
\,\,\,\,\,\,\,\,\,\,\,\,\,\, $g_{_X}^{N}$ \,\, &
\,\,\,\,\,\,\,\,\,\,\,\,\,\,
$g_{_{X}}^{C}$ \,\, \\
\hline
\,\,\,\,\,\,\,\,\,\,\, $ 0.5$ \,\,\,& \, $16.29 - 15.84$ & \, $12.64 - 12.40$ \,\,\, \\
\,\,\,\,\,\,\,\,\,\,\, $ 1.0 $ \,\,\,& \, $16.38 - 15.94$ & \, $12.68 - 12.44$ \,\,\, \\
\,\,\,\,\,\,\,\,\,\,\, $ 1.5$ \,\,\,& \, $16.46 - 16.04$ & \, $12.72 - 12.48$ \,\,\, \\
\hline \hline
\end{tabular}
\label{table:coupling}}
\end{table}
\begin{table}[h]
\tbl{The partial widths related to the data from the fitted
parameters $\phi$ and $\Lambda_{_X}$.} {\begin{tabular}{lllll}
\hline \hline $\Delta E$ & $\Gamma(X \to \gamma J/\psi)$ & $\Gamma(X
\to \gamma \psi(2S))$ & $\Gamma(X \to J/\psi\pi^+\pi^-\pi^0)$ &
$\Gamma(X \to J/\psi\pi^+\pi^-)$ \\
(MeV) & (KeV) & (KeV) & (KeV) & (KeV) \\
\hline
$ 0.5$ & $2.085 - 1.872$ & $5.066 - 6.525$ & $5.447 - 6.842$ & $5.648 - 6.612$ \\
$ 1.0$ & $2.082 - 1.864$ & $5.065 - 6.533$ & $5.253 - 6.606$ & $5.730 - 6.802$ \\
$ 1.5$ & $2.079 - 1.859$ & $5.058 - 6.537$ & $5.067 - 6.380$ & $5.801 - 6.977$ \\
\hline \hline
\end{tabular}
\label{table:widthdata} }
\end{table}
\begin{table}[h]
\tbl{Partial widths for $X(3872) \to \chi_{cJ} \pi^0$ decays.}
{\begin{tabular}{llll} \hline \hline $\Delta E$ \;\; & \;\;
$\Gamma(X \to \chi_{c0} \pi^0)$ \;\; & \hspace*{.2cm} $\Gamma(X \to
\chi_{c1} \pi^0)$\hspace*{.1cm} & \hspace*{.2cm}
$\Gamma(X \to \chi_{c2} \pi^0)$ \\
(MeV) \;\; & \;\;\;\;\;\; (KeV) \;\; & \;\;\;\;\;\; (KeV) \;\; &
\;\;\;\;\;\; (KeV) \\
\hline
$ 0.5 $ \,\,\,& \,\,\,\, $22.41 - 21.97$ & \,\,\,\, $0.294 - 0.276$ \,\,\,& \,\,\,\, $207.5 - 8.335$ \\
$ 1.0 $ \,\,\,& \,\,\,\, $22.60 - 22.40$ & \,\,\,\, $0.296 - 0.281$ \,\,\,& \,\,\,\, $211.8 - 8.618$ \\
$ 1.5 $ \,\,\,& \,\,\,\, $22.78 - 22.76$ & \,\,\,\, $0.296 - 0.285$ \,\,\,& \,\,\,\, $215.9 - 8.880$ \\
\hline \hline
\end{tabular}
\label{table:width1pi}  }
\end{table}

In Table.~\ref{table:coupling} we present our numerical results of
the coupling constants $g_{_X}^{N}$ and $g_{_X}^{C}$. Our bigger
results than the corresponding ones in the case $X(3872)$ with
$J^{PC} = 1^{++}$~\cite{Dong:2009yp} are because we need stronger
attractive interaction in the $P-$wave case to compensate the
repulsive interaction induced by angular momentum.

In terms of the isospin basis, one can write the wave function of
$X(3872)$ as
\begin{eqnarray}
|X(3872)\rangle = 0.988 \times |X(3872)\rangle_{I=0} + 0.156 \times
|X(3872)\rangle_{I=1}. \label{eq:wavex}
\end{eqnarray}
But this dominant isospin singlet component does not mean the decay
$X \to J/\psi \pi^+\pi^-$ is strongly suppressed compared to $X \to
J/\psi \pi^+\pi^-\pi^0$ decay because the later process is strongly
suppressed by the phase space factor. The explicit calculation
yields the ratio consistent with the experimental data given by
(\ref{eq:data3pto2p}).

In Table.~\ref{table:widthdata} we give the partial widths for
decays with $J/\psi$ or $\psi(2S)$ in the final states. All results
are of order of KeV. The inclusion of other components may change
our results. In case of $J/\psi\omega$ and $J/\psi\rho$ are the
components~\cite{Dong:2009yp,Swanson:2003tb}, the magnitudes of the
strong decay widths may be increased and the results depend on the
probability of these components. In this sense, if the strong decay
widths for tensor $X(3872)$ are observed bigger than our present
results, one may conclude that it cannot be a pure $DD^{\ast}$
molecule and other component should be included. If the $X(3872)$ is
regarded as a mixing state of $c\bar{c}$ and $DD^{\ast}$, one may
borrow the lesson from the $X(3872)$ with $1^{++}$
case~\cite{Dong:2008gb} to naively expect that this change of the
wave function of $X(3872)$ may increase the magnitude of radiative
decay width give in Table.~\ref{table:widthdata}. Although all these
cases should be calculated in detail, the precise measurement of the
strong decays can provide some clues on the structure of $X(3872)$.

Table.~\ref{table:width1pi} is the summary of our numerical results
for the decays of $X(3872) \to \chi_{cJ}\pi^0$. One can yield the
following ratio of the partial widths
\begin{eqnarray}
\Gamma(X \to \chi_{c0}\pi^0):\Gamma(X \to \chi_{c1}\pi^0) & \simeq &
1:0.013,\label{eq:ratiotensor}
\end{eqnarray}
which indicates that, compared with $X \to \chi_{c0}\pi^0$, $X \to
\chi_{c1}\pi^0$ is strongly suppressed.

Similarly to the radiative decay, including other components may
change our numerical results. In case of the $c\bar{c}$ and
$J/\psi\omega$ components are included, the magnitudes for the
partial widths might be changed but their ratio must be kept since
they are definitely isospin singlets so they do not contribute to
these decays. On the contrary, complication arises if $X(3872)$ has
a $J/\psi\rho$ component. So that, in the case $X(3872)$ with
$J^{PC} = 2^{-+}$, the strong suppression of the decay $X(3872) \to
\chi_{c1}\pi^0$ compared with the decay $X(3872) \to \chi_{c0}\pi^0$
may signal the pure $DD^{\ast}$ molecular structure of $X(3872)$.

\section{Conclusion}

By regarding the hidden charm meson $X(3872)$ with $J^{PC} = 2^{-+}$
as a $DD^{\ast}$ bound state, we found that our model with dominant
isospin zero component can explain the existing date quite well. We
also predicted the partial widths for $X(3872) \to \gamma J/\psi$,
$X(3872) \to \gamma\psi(2S)$, $X(3872) \to J/\psi \pi^+\pi^-$,
$X(3872) \to J/\psi \pi^+\pi^-\pi^0$ and $X(3872) \to
\chi_{cJ}\pi^0$. And, from our study, comparison of these values
with the future experiments will shed a light on the nature of
$X(3872)$.


\section*{Acknowledgments}

This work is supported in part by Grant-in-Aid for Scientific
Research on Innovative Areas (No. 2104) ``Quest on New Hadrons with
Variety of Flavors'' from MEXT. The work of M.H. is supported in
part by the Grant-in-Aid for Nagoya University Global COE Program
``Quest for Fundamental Principles in the Universe: from Particles
to the Solar System and the Cosmos'' from MEXT, the JSPS
Grant-in-Aid for Scientific Research (S) $\sharp$ 22224003, (c)
$\sharp$ 20540262. The work of Y.M. is supported in part by the
National Science Foundation of China (NNSFC) under grant No.
10905060.









\begin{thebibliography}{0}    

\bibitem{Harada:2010bs}
  M.~Harada and Y.~L.~Ma,
  arXiv:1010.3607 [hep-ph].


\bibitem{al.:2010jr}
  P.~d.~A.~al.  [BABAR Collaboration],
  arXiv:1005.5190 [hep-ex].


\bibitem{Weinberg:1962hj}
  S.~Weinberg,
  Phys.\ Rev.\  {\bf 130}, 776 (1963).

\bibitem{Salam:1962ap}
  A.~Salam,
  Nuovo Cim.\  {\bf 25} (1962) 224.


\bibitem{Amsler:2008zz}
  C.~Amsler {\it et al.}  [Particle Data Group],
  Phys.\ Lett.\  B {\bf 667}, 1 (2008).


\bibitem{Abe:2005ix}
  K.~Abe {\it et al.}  [Belle Collaboration],
  arXiv:hep-ex/0505037.


\bibitem{:2008rn}
  B.~Aubert {\it et al.}  [BABAR Collaboration],
  Phys.\ Rev.\ Lett.\  {\bf 102}, 132001 (2009)
  [arXiv:0809.0042 [hep-ex]].

\bibitem{Dong:2009yp}
  Y.~Dong, A.~Faessler, T.~Gutsche, S.~Kovalenko and V.~E.~Lyubovitskij,
  Phys.\ Rev.\  D {\bf 79}, 094013 (2009)
  [arXiv:0903.5416 [hep-ph]].


\bibitem{Swanson:2003tb}
  E.~S.~Swanson,
  Phys.\ Lett.\  B {\bf 588}, 189 (2004)
  [arXiv:hep-ph/0311229].


\bibitem{Dong:2008gb}
  Y.~b.~Dong, A.~Faessler, T.~Gutsche and V.~E.~Lyubovitskij,
  Phys.\ Rev.\  D {\bf 77}, 094013 (2008)
  [arXiv:0802.3610 [hep-ph]].


\end{thebibliography}
\end{document}